\documentclass[a4paper, oneside, twocolumn, notitlepage, 10pt]{extarticle_ecoc}
\usepackage{ecoc}
\usepackage[helvet]{sfmath}

\usepackage[binary-units, per-mode=symbol]{siunitx}
\DeclareSIUnit{\sample}{Sa}
\DeclareSIUnit{\baud}{Bd}
\DeclareSIUnit{\bit}{bit}
\DeclareSIUnit{\fourd}{4D}
\DeclareSIUnit{\eightd}{8D}
\DeclareSIUnit{\dBm}{dBm}
\DeclareSIUnit{\dB}{dB}
\DeclareSIUnit{\bps}{bps}

\usepackage{glossaries}
\glsdisablehyper
\newcommand{\SetCapsType}{normalcaps}

\makeatletter
\providecommand{\SetCapsType}{smallcaps}

\long\def\@scTrue{smallcaps}
\long\def\@scFalse{normalcaps}
\newcommand{\acroSCaps}[1]{%
 \begingroup
  \ifx\SetCapsType\@scTrue 
    \textsc{#1}%
  \else
    \MakeUppercase{#1}%
  \fi
  \endgroup
}
\makeatother

\newcommand{\nAcronym}[4][]{%
	\newacronym[#1]{#2}{#3}{#4}
}

\makeatletter
\@ifpackageloaded{babel}{%
    \newcommand{\usuk}[2]{%
        \iflanguage{USenglish}{#1}{#2}%
    }%
}{%
    \newcommand{\usuk}[2]{%
        #1%
    }%
}%
\makeatother

\usepackage[shortcuts]{extdash}
\nAcronym{2A8PSK}{\acroSCaps{2a8psk}}{2-ary amplitude 8-ary phaseshift keying}

\nAcronym{3CCMCF}{\acroSCaps{3cc-mcf}}{3-core coupled-core multi-core fiber}

\nAcronym{4D}{\acroSCaps{4d}}{four-dimensional}
\nAcronym{4D64PRS}{\acroSCaps{4d-64prs}}{\usuk{four-dimensional 64-ary polarization-ring-switching}{four-dimensional 64-ary polarisation-ring-switching}}

\nAcronym{5B4D2A8PSK}{\acroSCaps{5b4d-2a8psk}}{5-bit four-dimensional two-amplitude 8-ary phase-shift keying}

\nAcronym{6B4D2A8PSK}{\acroSCaps{6b4d-2a8psk}}{6-bit four-dimensional two-amplitude 8-ary phase-shift keying}

\nAcronym{8D}{\acroSCaps{8d}}{eight-dimensional}\nAcronym{8D2048PRS}{\acroSCaps{8d-2048prs}}{eight-dimensional 2048-ary polarization-ring-switching}
\nAcronym{8D2048PRST1}{\acroSCaps{8d-2048prs-t1}}{eight-dimensional 2048-ary polarization-ring-switching type 1}
\nAcronym{8D2048PRST2}{\acroSCaps{8d-2048prs-t2}}{eight-dimensional 2048-ary polarization-ring-switching type 2}

\nAcronym{ABC}{\acroSCaps{abc}}{automatic bias controller}
\nAcronym{ADC}{\acroSCaps{adc}}{analog-to-digital converter}
\nAcronym{AIR}{\acroSCaps{air}}{achievable information rate}
\nAcronym{AOM}{\acroSCaps{aom}}{acoustic optical modulator}
\nAcronym{AR}{\acroSCaps{ar}}{achievable rate}
\nAcronym{ASE}{\acroSCaps{ase}}{amplified spontaneous emission}
\nAcronym{ASIC}{\acroSCaps{asic}}{application-specific integrated circuit}
\nAcronym{AWGN}{\acroSCaps{awgn}}{additive white Gaussian noise}

\nAcronym{BER}{\acroSCaps{ber}}{bit error rate}
\nAcronym{BICM}{\acroSCaps{bicm}}{bit-interleaved coded modulation}
\nAcronym{BMD}{\acroSCaps{bmd}}{bit-metric decoding}
\nAcronym{BPD}{\acroSCaps{bpd}}{balanced photo-diode}
\nAcronym{BPS}{\acroSCaps{bps}}{blind phase search}
\nAcronym{BRGC}{\acroSCaps{brgc}}{binary reflected Gray code}
\nAcronym{BTB}{\acroSCaps{btb}}{back-to-back}

\nAcronym{CCDM}{\acroSCaps{ccdm}}{constant composition distribution matching}
\nAcronym{CCF}{\acroSCaps{ccf}}{\usuk{coupled-core fiber}{coupled-core fibre}}
\nAcronym{CD}{\acroSCaps{cd}}{chromatic dispersion}
\nAcronym{CIR}{\acroSCaps{cir}}{channel impulse response}
\nAcronym{CMUX}{\acroSCaps{cmux}}{core multiplexer}
\nAcronym{ChUT}{\acroSCaps{chut}}{channel under test}
\nAcronym{CUT}{\acroSCaps{cut}}{channel under test}
\nAcronym{CPE}{\acroSCaps{cpe}}{carrier phase estimation}
\nAcronym{CAGR}{\acroSCaps{cagr}}{compound annual growth rate}

\nAcronym{DA}{\acroSCaps{da}}{driver amplifier}
\nAcronym{DAC}{\acroSCaps{dac}}{digital-to-analog converter}
\nAcronym{DCF}{\acroSCaps{dcf}}{\usuk{dispersion compensated fiber}{dispersion compensated fibre}}
\nAcronym{DCI}{\acroSCaps{dci}}{data center interconnect}
\nAcronym{DFB}{\acroSCaps{dfb}}{distributed feedback}
\nAcronym{DGD}{\acroSCaps{dgd}}{differential group delay}
\nAcronym{DM}{\acroSCaps{dm}}{distribution matcher}
\nAcronym{DMGD}{\acroSCaps{dmgd}}{differential mode group delay}
\nAcronym{DP}{\acroSCaps{dp}}{\usuk{dual-polarization}{dual-polarisation}}
\nAcronym{DPC}{\acroSCaps{dpc}}{digital pre-compensation}
\nAcronym{DPE}{\acroSCaps{dpe}}{digital pre-emphasis}
\nAcronym{DPIQ}{\acroSCaps{dp-iqm}}{\usuk{dual-polarization IQ-modulator}{dual-polarisation IQ-modulator}}
\nAcronym{DRE}{\acroSCaps{dre}}{digital resolution enhancer}
\nAcronym{DSO}{\acroSCaps{dso}}{digital sampling oscilloscope}
\nAcronym{DSP}{\acroSCaps{dsp}}{digital signal processing}
\nAcronym{DUT}{\acroSCaps{dut}}{device-under-test}
\nAcronym{DWDM}{\acroSCaps{dwdm}}{dense wavelength-division multiplexing}

\nAcronym{ECL}{\acroSCaps{ecl}}{external cavity laser}
\nAcronym{ED}{\acroSCaps{ed}}{Eucledian distance}
\nAcronym{EDFA}{\acroSCaps{edfa}}{\usuk{erbium-doped fiber amplifier}{erbium-doped fibre amplifier}}
\nAcronym{ENOB}{\acroSCaps{enob}}{effective number of bits}
\nAcronym{ESS}{\acroSCaps{ess}}{enumerative sphere shaping}

\nAcronym{FDE}{\acroSCaps{fde}}{\usuk{frequency domain equalizer}{frequency domain equaliser}}
\nAcronym{FEC}{\acroSCaps{fec}}{forward error correction}
\nAcronym{FFT}{\acroSCaps{fft}}{fast Fourier transform}
\nAcronym{FMF}{\acroSCaps{fmf}}{\usuk{few-mode fiber}{few-mode fibre}}
\nAcronym[plural=FM-MCF, firstplural=\usuk{few-mode multi-core fibers}{few-mode multi-core fibres}]{FM-MCF}{\acroSCaps{fm-mcf}}{\usuk{few-mode multi-core fiber}{few-mode multi-core fibre}}
\nAcronym{FWM}{\acroSCaps{fwm}}{four-wave mixing}

\nAcronym{GFF}{\acroSCaps{gff}}{gain flattening filter}
\nAcronym{GIMMF}{\acroSCaps{gi-mmf}}{\usuk{graded-index multi-mode fiber}{graded-index multi-mode fibre}}
\nAcronym{GMI}{\acroSCaps{gmi}}{\usuk{generalized mutual information}{generalised mutual information}}
\nAcronym{GS}{\acroSCaps{gs}}{geometric shaping}
\nAcronym{GV}{\acroSCaps{gv}}{group-velocity}
\nAcronym{GVD}{\acroSCaps{gvd}}{group-velocity dispersion}

\nAcronym{HDFEC}{\acroSCaps{hd-fec}}{hard decision forward error correction}

\nAcronym[plural=IL, firstplural=insertion losses (\textsc{il})]{IL}{\acroSCaps{il}}{insertion loss}
\nAcronym{IFFT}{\acroSCaps{ifft}}{inverse fast Fourier transform}
\nAcronym{ISI}{\acroSCaps{isi}}{intersymbol interference}

\nAcronym{KK}{\acroSCaps{kk}}{Kramers-Kronig}

\nAcronym{LCOS}{\acroSCaps{LCoS}}{liquid crystal on silicon}
\nAcronym{LDPC}{\acroSCaps{ldpc}}{low-density parity-check}
\nAcronym{LEAF}{\acroSCaps{leaf}}{\usuk{large effective area fiber}{large effective area fibre}}
\nAcronym{LMS}{\acroSCaps{lmf}}{least means square}
\nAcronym{LLR}{\acroSCaps{llr}}{log-likelihood ratio}
\nAcronym{LO}{\acroSCaps{lo}}{local oscillator}
\nAcronym{LP}{\acroSCaps{lp}}{\usuk{linear polarized}{linear polarised}}
\nAcronym{LSPS}{\acroSCaps{lsps}}{\usuk{loop-synchronized polarization scrambler}{loop-synchronised polarisation scrambler}}
\nAcronym{LUT}{\acroSCaps{lut}}{lookup table}

\nAcronym{MB}{\acroSCaps{mb}}{Maxwell-Bolzmann}
\nAcronym{MCF}{\acroSCaps{mcf}}{\usuk{multi-core fiber}{multi-core fibre}}
\nAcronym{MDG}{\acroSCaps{mdg}}{mode dependent gain}
\nAcronym[plural=MDL, firstplural=mode-dependent losses (\textsc{mdl})]{MDL}{\acroSCaps{mdl}}{mode-dependent loss}
\nAcronym{MDM}{\acroSCaps{mdm}}{mode division multiplexing}
\nAcronym{MF}{\acroSCaps{mf}}{matched filter}
\nAcronym{MI}{\acroSCaps{mi}}{mutual information}
\nAcronym{MIMO}{\acroSCaps{mimo}}{multiple-input multiple-output}
\nAcronym{MMA}{\acroSCaps{mma}}{multi-modulus algorithm}
\nAcronym{MMF}{\acroSCaps{mmf}}{\usuk{multi-mode fiber}{multi-mode fibre}}
\nAcronym{MMSE}{\acroSCaps{mmse}}{minimum mean square error}
\nAcronym{MPLC}{\acroSCaps{mplc}}{multi-plane light converter}
\nAcronym{MSE}{\acroSCaps{mse}}{mean squared error}
\nAcronym{MUX}{\acroSCaps{mux}}{multiplexer}
\nAcronym{MZM}{\acroSCaps{mzm}}{Mach-Zehnder modulator}

\nAcronym{NF}{\acroSCaps{nf}}{noise figure}
\nAcronym{NGMI}{\acroSCaps{ngmi}}{\usuk{normalized generalized mutual information}{normalised generalised mutual information}}
\nAcronym{NIR}{\acroSCaps{nir}}{near infra-red}

\nAcronym{OBTB}{\acroSCaps{obtb}}{optical back-to-back}
\nAcronym{ODL}{\acroSCaps{odl}}{optical delay line}
\nAcronym{OFDR}{\acroSCaps{ofdr}}{optical frequency-domain reflectometer}
\nAcronym{OFDM}{\acroSCaps{ofdm}}{orthogonal frequency division multiplexing}
\nAcronym{OMFT}{\acroSCaps{omft}}{optical multi-format transmitter}
\nAcronym{OSA}{\acroSCaps{osa}}{\usuk{optical spectrum analyzer}{optical spectrum analyser}}
\nAcronym{OSNR}{\acroSCaps{osnr}}{optical signal-to-noise ratio}
\nAcronym{OTDR}{\acroSCaps{otdr}}{optical time-domain reflectometer}
\nAcronym{OTF}{\acroSCaps{otf}}{optical tunable filter}
\nAcronym{OVNA}{\acroSCaps{ovna}}{\usuk{optical vector network analyzer}{optical vector network analyser}}

\nAcronym{PAM}{\acroSCaps{pam}}{pulse-amplitude modulation}
\nAcronym{PAS}{\acroSCaps{pas}}{probabilistic amplitude shaping}
\nAcronym{PAPR}{\acroSCaps{papr}}{peak-to-avarage power ratio}
\nAcronym{PBS}{\acroSCaps{pbs}}{polarization beam splitter}
\nAcronym{PCVD}{\acroSCaps{pcvd}}{plasma chemical vapor depostion}
\nAcronym{PDL}{\acroSCaps{pdl}}{polarization-dependent loss}
\nAcronym{PDM}{\acroSCaps{pdm}}{\usuk{polarization-division multiplexing}{polarisation-division multiplexing}}
\nAcronym{PL}{\acroSCaps{pl}}{photonic lantern}
\nAcronym{PMBPSK}{\acroSCaps{pm-bpsk}}{polarization-multiplexed binary phase-shift-keying}
\nAcronym{PMQPSK}{\acroSCaps{pm-qpsk}}{polarization-multiplexed quaternary phase-shift-keying}
\nAcronym{PM8QAM}{\acroSCaps{pm-8qam}}{polarization-multiplexed 8-ary quadrature amplitude modulation}
\nAcronym{PMD}{\acroSCaps{pmd}}{polarization mode dispersion}
\nAcronym{PNOB}{\acroSCaps{pnob}}{physical number of bits}
\nAcronym{PRBS}{\acroSCaps{prbs}}{pseudorandom bit sequence}
\nAcronym{PS}{\acroSCaps{ps}}{probabilistic shaping}

\nAcronym{QAM}{\acroSCaps{qam}}{quadrature amplitude modulation}
\nAcronym{QPSK}{\acroSCaps{qpsk}}{quaternary phase-shift-keying}

\nAcronym{RRC}{\acroSCaps{rrc}}{root-raised-cosine}
\nAcronym{RF}{\acroSCaps{rf}}{radio frequency}
\nAcronym{SamPerSym}{\acroSCaps{sps}}{samples per symbol}
\nAcronym{SDFEC}{\acroSCaps{sd-fec}}{soft decision forward error correction}
\nAcronym{SDM}{\acroSCaps{sdm}}{space-division multiplexing}
\nAcronym{SE}{\acroSCaps{se}}{spectral efficiency}
\nAcronym{SER}{\acroSCaps{ser}}{symbol error rate}
\nAcronym{SA}{\acroSCaps{sa}}{simulated annealing}
\nAcronym{SLM}{\acroSCaps{slm}}{spatial light modulator}
\nAcronym{SSFM}{\acroSCaps{ssfm}}{split-step Fourier method}
\nAcronym{SMF}{\acroSCaps{smf}}{\usuk{single-mode fiber}{single-mode fibre}}
\nAcronym{SNR}{\acroSCaps{snr}}{signal-to-noise ratio}
\nAcronym[firstplural=\usuk{states of polarization (\acroSCaps{sop})}{states of polarisation (\acroSCaps{sop})}]{SOP}{\acroSCaps{sop}}{\usuk{state of polarization}{state of polarisation}}
\nAcronym{SPM}{\acroSCaps{spm}}{self-phase modulation}
\nAcronym{SPS}{\acroSCaps{sps}}{\usuk{synchronized polarization scrambler}{synchronised polarisation scrambler}}
\nAcronym{SSMF}{\acroSCaps{ssmf}}{\usuk{standard single-mode fiber}{standard single-mode fibre}}
\nAcronym{SVD}{\acroSCaps{svd}}{singular value decomposition}
\nAcronym{SWI}{\acroSCaps{swi}}{swept wavelength interferometry}

\nAcronym{TDE}{\acroSCaps{tde}}{\usuk{time domain equalizer}{time domain equaliser}}
\nAcronym{TDMSDM}{\acroSCaps{tdm-sdm}}{time-domain multiplexed space-division multiplexing}
\nAcronym{TE}{\acroSCaps{TE}}{transverse electric}
\nAcronym{TM}{\acroSCaps{TM}}{transverse magnetic}
\nAcronym{TH4D}{\acroSCaps{th-4d}}{time domain hybrid four-dimensional}
\nAcronym{TH4D2A8PSK}{\acroSCaps{th-4d-2a8psk}}{time domain hybrid four-dimensional two-amplitude eight-phase shift keying}

\nAcronym{VOA}{\acroSCaps{voa}}{variable optical attenuator}

\nAcronym{WDM}{\acroSCaps{wdm}}{wavelength-division multiplexing}
\nAcronym{WSS}{\acroSCaps{wss}}{wavelength selective switch}

\nAcronym{XPM}{\acroSCaps{xpm}}{cross-phase modulation}
\nAcronym{XT}{\acroSCaps{xt}}{cross-talk}

\nAcronym{3DWG}{\acroSCaps{3dwg}}{3D-waveguide} 

\usepackage{subcaption}

\newcommand{\LP}[1]{LP\textsubscript{#1}}

\usepackage{fancyhdr}

\fancypagestyle{firststyle}
{
  \fancyhf{}
  \fancyfoot[C]{\footnotesize \copyright~IEEE 2020}
}

\usepackage[capitalise]{cleveref}

\begin{document}
    \selectlanguage{english}    


    \title{Experimental validation of MDL emulation and estimation techniques for SDM transmission systems}%


    
    \author{
        Menno~van~den~Hout\textsuperscript{(1,*)},
        Ruby~S.~B.~Ospina\textsuperscript{(1,2)},
        Sjoerd~van~der~Heide\textsuperscript{(1)},
        Juan~Carlos~Alvarado-Zacarias\textsuperscript{(3)},\\
        Jose~Enrique~Antonio-L\'opez\textsuperscript{(3)},
        Marianne~Bigot-Astruc\textsuperscript{(4)},
        Adrian~Amezcua~Correa\textsuperscript{(4)},\\
        Pierre~Sillard\textsuperscript{(4)},
        Rodrigo~Amezcua-Correa\textsuperscript{(3)},
        Darli~A.~A.~Mello\textsuperscript{(2)}
        and Chigo~Okonkwo\textsuperscript{(1)}\vspace{-1mm}
    }

    \maketitle                  


    \begin{strip}
        \begin{author_descr}

            \textsuperscript{(1)} High Capacity Optical Transmission Laboratory,
            Electro-Optical Communications Group,\\
            Eindhoven University of Technology, the Netherlands,
            \textsuperscript{(*)}{\uline{m.v.d.hout@tue.nl}}

            \textsuperscript{(2)} DECOM, School of Electrical and Computer Engineering,  University of Campinas, Brazil

            \textsuperscript{(3)} CREOL, The College of Optics and Photonics, University of Central Florida, USA

            \textsuperscript{(4)} Prysmian Group, 644 Boulevard Est, Billy Berclau, 62092 Haisnes Cedex, France
            \vspace{-2mm}
        \end{author_descr}
    \end{strip}

    \setstretch{1.1}


    \begin{strip}
        \begin{ecoc_abstract} %
            We experimentally validate a \acrfull{MDL} estimation technique employing a correction factor to remove the MDL estimation dependence on the SNR when using a \acrfull{MMSE} equalizer. A reduction of the MDL estimation error is observed for both transmitter-side and in-span MDL emulation.\vspace{-3mm}
        \end{ecoc_abstract}
    \end{strip}

\thispagestyle{firststyle}
    \section{Introduction}\vspace{-2mm}
    \Gls{SDM} provides a significant capacity increase over the use of \gls{SMF}, by transmitting over multiple modes or cores in a single fiber.\cite{winzer2014optical}
    \Gls{MDL} and \gls{MDG} are known to limit the performance of \gls{SDM} transmission and can even cause system outage \cite{ho2011mode}\cite{Mello:20}. MDL/MDG are generated by unequal attenuation and/or amplification of the guided modes in amplifiers, spatial (de)multiplexers, switches, fibers, connectors and splices. 
    
    At a component level, \gls{MDL} can be characterized by means of an \gls{OVNA}\cite{Rommel:2017,weerdenburg2019} or digital holography techniques\cite{MazurDH,JC_DH}, while at a system level, the \gls{MDL} is usually estimated from the channel transfer function obtained by the \gls{MIMO} equalizer.
    In \parencite{ospina2020dsp}, we showed that the \gls{MDL} estimation based on \gls{MMSE} \gls{MIMO} equalizers depends on the \gls{SNR}, resulting in an underestimation of the \gls{MDL} at low \gls{SNR}. We also proposed a correction factor that improves the estimation process for moderate levels of \gls{MDL}.

\begin{figure}[b]
    	\centering
    	\vspace*{-6mm}
    	\begin{subfigure}{0.75\linewidth}
    		\centering
    		\includegraphics[width=\linewidth]{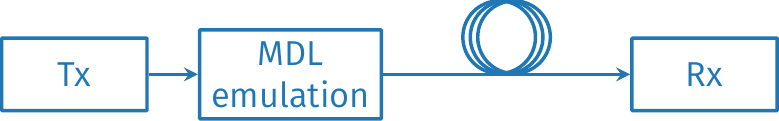}
    		\caption{}
    		\label{fig:mdl_tx_emul}
    		\vspace{3mm}
    	\end{subfigure}
    	\begin{subfigure}{0.75\linewidth}
    		\centering
    		\includegraphics[width=\linewidth]{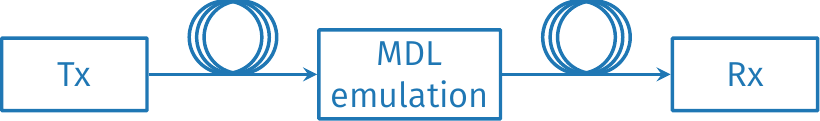}
    		\caption{}
    		\label{fig:mdl_span_emu}
    		\vspace{3mm}
    	\end{subfigure}
    	\caption{Schematic representation of \gls{MDL} emulation at the transmitter side (a), and in-span \gls{MDL} emulation (b).}
    	\label{fig:mdl_emulation_schemes}
    \end{figure}

    \Gls{MDL} can be artificially introduced by purposely attenuating or amplifying the fiber modes by different factors. For example, by using multi-mode amplifiers with an inherent \gls{MDG} or by changing the powers of the single mode tributaries of mode (de)multiplexers using amplifiers or attenuators. As a benefit, the latter allows controlling the powers of the modes individually and hence to vary the induced \gls{MDL}, which is not possible by using a multi-mode amplifier as there is no control over the individual gains of the modes. 
    \Gls{MDL} emulation by varying individual mode powers can be done by placing an \gls{MDL} emulator stage directly after the transmitter\cite{MizunoMDL2015} (\cref{fig:mdl_tx_emul}), or in-span (\cref{fig:mdl_span_emu}). Compared to emulation at the transmitter side, in-span configuration enables a more approximate emulation of the \gls{MDL} introduced by \gls{SDM} components like multi-mode amplifiers, optical switches, spatial (de)multiplexers and \gls{MMF}, as the powers of the modes are varied after the modes have mixed.
    
    
    The correction factor that improves the \gls{MDL} estimation was initially proposed in \parencite{ospina2020dsp} and experimentally validated in {\parencite{Ospina2020Preprint}} for a back-to-back and single span transmission scheme using an \gls{MDL} emulator at the transmitter side. In this work, we extend the experimental validation and show that the correction factor reduces the \gls{MDL} estimation error both for the case of \gls{MDL} emulation at the transmitter side, as well as for the in-span \gls{MDL} emulation scenario.
    
    \begin{figure*}[!t]
        \centering
        \vspace{-5mm}
        \includegraphics[width=\linewidth]{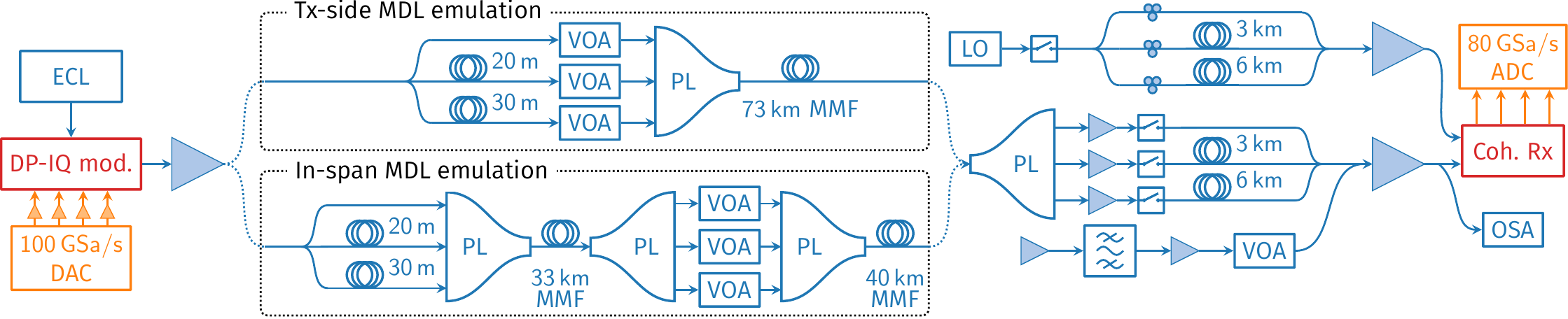}
        \vspace*{-4mm}
        \caption{Experimental setup for MDL emulation in 3-mode transmission. The transmitter generates 16-QAM symbols at 25 GBd, which are subsequently split and delayed to create the input tributaries for the \gls{PL}. \Glspl{VOA} are placed directly after the transmitter or in the fiber span in order to emulate MDL. The multi-mode signal is transmitted over in total \SI{73}{\km} of \gls{MMF} \cite{sillard201650}. At the receiver, a \gls{TDMSDM} scheme is employed, and a noise-loading stage is used to vary the \gls{OSNR}.}%
        \glsreset{VOA}
        \glsreset{OSNR}
        \label{fig:setup}
    \end{figure*}
    
    \vspace*{-1mm}
    \section{MDL estimation correction factor}
    The peak-to-peak \gls{MDL} of a transmission link can be computed from the ratio between the maximum and minimum eigenvalues \(\lambda_i^2\) of the operator $\mathbf{H}\mathbf{H}^{H}$, where $\mathbf{H}$ is the channel transfer matrix and $(.)^{H}$ denotes the Hermitian transpose operator\cite{ho2011mode}. The relation between the channel transfer matrix and the \gls{MMSE} equalizer transfer matrix depends on the \gls{SNR}, then, the eigenvalues obtained from the equalizer \(\lambda^{2}_{i_{\mathrm{MMSE}}}\), are related to the actual eigenvalues \(\lambda_i^2\), as\cite{ospina2020dsp}
    \begin{equation}
        \lambda^{2}_{i_{\mathrm{MMSE}}} = \left[\frac{\left(\lambda^{2}_{i}\right)^{-1}}{\mathrm{SNR}^{2}} + \frac{2}{\mathrm{SNR}} + \lambda^{2}_{i} \right].
        \label{Eq:Lambdarelation}
    \end{equation}
    For a known SNR, this relation can be inverted, resulting in
    \begin{equation}
        \label{Eq:roots}
        \resizebox{1\hsize}{!}{$%
        \lambda^{2}_{i}= \frac{ \left[ \mathrm{SNR}^{2}\,\lambda^{2}_{i_{\mathrm{MMSE}}}-2\,\mathrm{SNR} \right] {\pm} \sqrt{\left[\mathrm{SNR}^{2}\,\lambda^{2}_{i_{\mathrm{MMSE}}}-2\,\mathrm{SNR} \right]^2 - 4\,\mathrm{SNR}^{2}} }{2\,\mathrm{SNR}^{2}},
        $}
    \end{equation}
    where the positive solution of \eqref{Eq:roots} is a correction factor proposed to recover \(\lambda_i^2\)~\,\cite{ospina2020dsp} and improve the estimation for moderate levels of \gls{MDL}.
    
    \vspace*{-2mm}
    \section{Experimental setup}
    
     \label{sec:setup}
    The experimental setup used for MDL emulation is depicted in \cref{fig:setup}. At the transmitter, a \gls{PRBS} of 2\textsuperscript{16} polarization-multiplexed 16-QAM symbols is generated at \SI{25}{\giga\baud}. Pulse shaping at the transmitter is done using a \gls{RRC} filter with $\beta = 0.01$. The shaped signal is converted to the analog domain by a \SI{100}{\giga\sample\per\second} \gls{DAC} followed by RF\=/amplifiers. The resulting signal modulates the output of an \gls{ECL} operating at \SI{193.4}{\tera\hertz} using a dual-polarization in-phase and quadrature modulator. After modulation, the signal is amplified by an \gls{EDFA}, split and delayed by \SIlist{0; 20; 30}{\meter} to generate three decorrelated data streams that are multiplexed by a mode-selective \glsfirst{PL} \cite{velazquez2018scaling}. 
     The output of the \gls{PL} is connected to a \SI{73}{\km} fiber link consisting of 16 spools of \SI{50}{\micro\meter} core diameter graded-index \gls{MMF} \cite{sillard201650} with lengths varying from \SI{1.2}{\km} to \SI{8.9}{\km}. At the receiver side, a mode-selective \gls{PL} is used as mode de-multiplexer. 
  \begin{figure}[b]
  \vspace*{-6mm}
        \centering
        \includegraphics[width=0.94\linewidth]{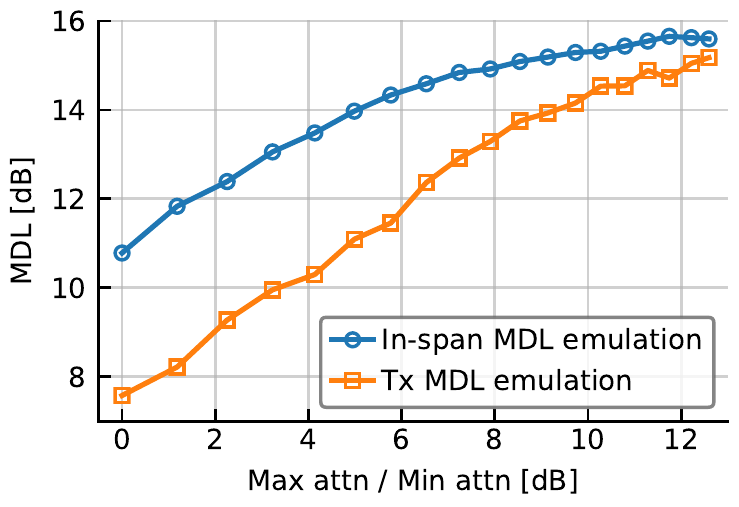}
        \caption{\Gls{MDL} versus attenuation ratio between the \LP{01} and \LP{11} modes. No noise loading is used.}%
        \vspace*{-6mm}
        \label{fig:attn_vs_mdl}
    \end{figure}
    
    In order to emulate \gls{MDL}, two different schemes are employed (see \cref{fig:setup}). For transmitter side \gls{MDL} emulation, \glspl{VOA} are placed between the decorrelation fibers and the inputs of the transmitter \gls{PL}. In-span \gls{MDL} emulation is achieved by placing two mode-selective \glspl{PL} connected by \glspl{VOA} after \SI{33}{\km} of \gls{MMF}.
    
     The receiver employs a time domain multiplexed (TDM)-SDM scheme \cite{RoyTDMSDM} that delays two flows by 3 km and 6 km of \gls{SMF} to reduce the required amount of coherent receivers. After the TDM-SDM stage, a noise-loading stage composed of two \glspl{EDFA}, a \gls{WSS} and a \gls{VOA} is placed to vary the \gls{OSNR} at the coherent receiver input. This noise-loading setup places a \SI{250}{\giga\hertz} wide noise-band around the \SI{193.4}{\tera\hertz} carrier. The average OSNR is measured by an \gls{OSA} after the last amplification stage. The average SNR at the receiver input is computed as  ${SNR= OSNR \, (T_{s} \times 12.5 \,\mathrm{GHz})}$ where $T_{s}$ is the symbol time \cite{essiambre2010capacity}. The noisy signal is amplified and converted from the optical to the electrical domain by the receiver front-end that integrates a second ECL as local oscillator (LO). The TDM electric signals are fed into \SI{80}{\giga\sample\per\second} \glspl{ADC} to be digitized. In the DSP block, the TDM streams are parallelized and down-sampled to two samples per symbol. Next, dispersion is digitally compensated and frequency offset is estimated and compensated for. The signal is matched-filtered by a RRC filter, and, finally, fully supervised \gls{MMSE} equalization is applied and \gls{MDL} is calculated from the \gls{MIMO} equalizer taps.


    \section{Results}\label{sec:results}

    First, the ability of the two \gls{MDL} emulation schemes to introduce \gls{MDL} is verified by varying the attenuation of the \glspl{VOA} and estimating the \gls{MDL}. In order to keep the launch power constant, the three \glspl{VOA} are initialized at an attenuation of \SI{-5}{\dB} and the attenuation of the \LP{11} modes is gradually increased, while decreasing the attenuation of the \LP{01} mode. In \cref{fig:attn_vs_mdl}, the estimated \gls{MDL} for different attenuation ratios between the \LP{01} and \LP{11} modes is shown, indicating the capability of \gls{MDL} emulation for both schemes. Due to the two extra \glspl{PL} in the transmission link for the in-span \gls{MDL} emulation compared to the transmitter side \gls{MDL} emulation, the initial \gls{MDL} is about \SI{2.5}{\dB} higher for the in-span scheme.

    \begin{figure}
        \captionsetup[subfigure]{labelformat=empty}
        \centering
        \subfloat[\label{fig:mdl_wo_corr_tx}]
        {
            \includegraphics[width=0.95\linewidth]{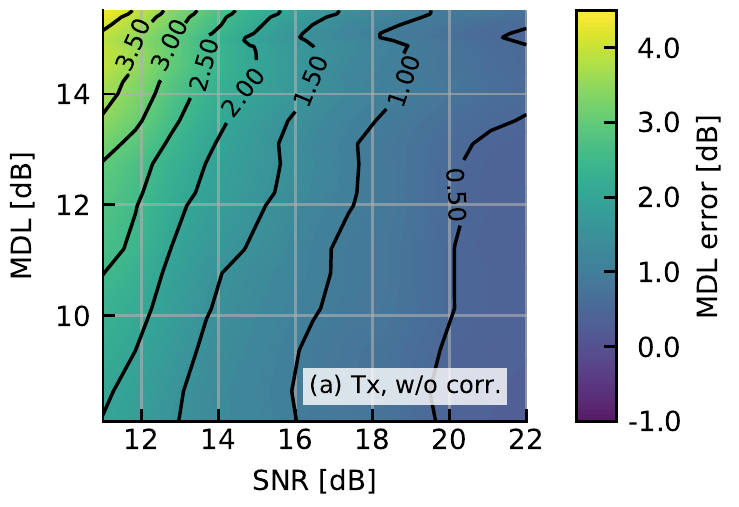}
        }\\ \vspace*{-3mm}
        \subfloat[\label{fig:mdl_w_corr_tx}]
        {
            \includegraphics[width=0.95\linewidth]{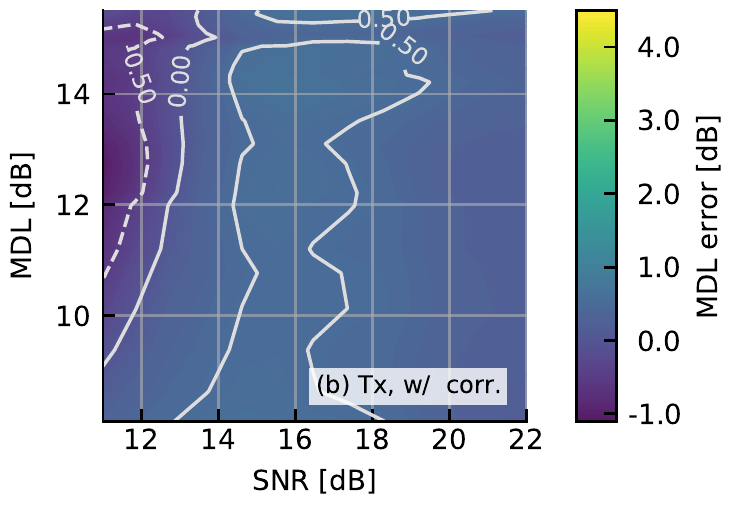}
        }\vspace*{-3mm}
        \caption{\Gls{MDL} estimation error without (a) and with (b) correction, as a function of the actual \gls{MDL} and the \gls{SNR}. The \gls{MDL} is emulated at the transmitter side.}
        \label{fig:mdl_tx}
    \end{figure}

\begin{figure}
        \captionsetup[subfigure]{labelformat=empty}
        \centering
        \subfloat[\label{fig:mdl_wo_corr_span}]
        {
            \includegraphics[width=0.95\linewidth]{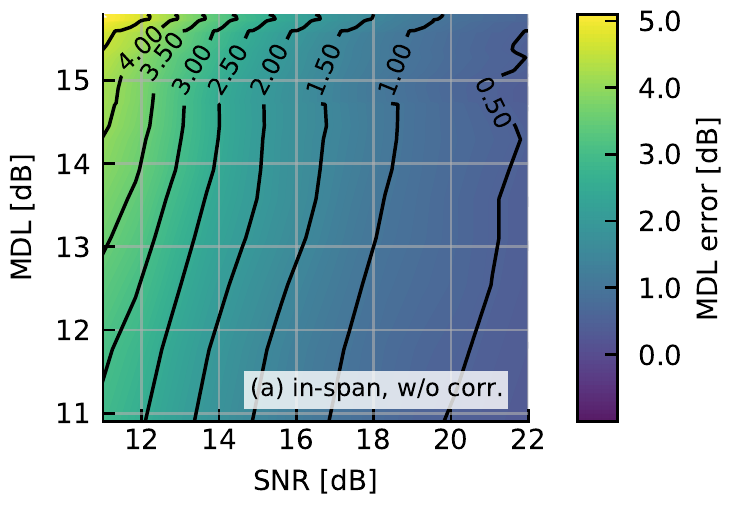}
        }\\  \vspace*{-3mm}
        \subfloat[\label{fig:mdl_w_corr_span}]
        {
            \includegraphics[width=0.95\linewidth]{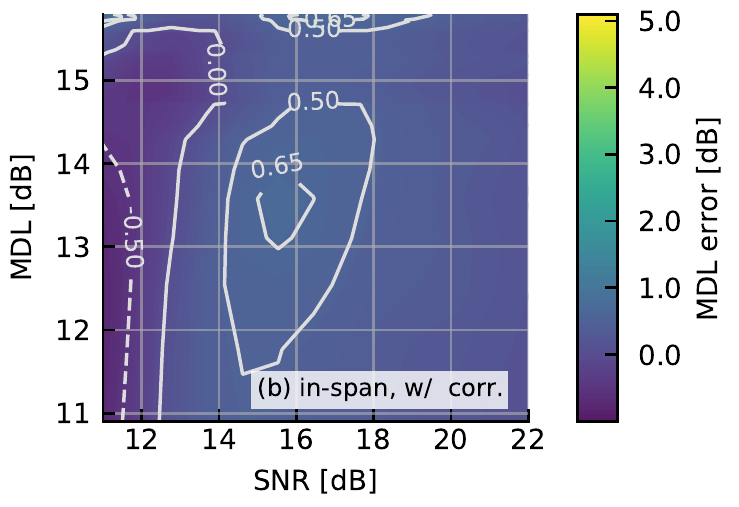}
        } \vspace*{-3mm}
        \caption{\Gls{MDL} estimation error without (a) and with (b) correction, as a function of the actual \gls{MDL} and the \gls{SNR}. The \gls{MDL} is emulated at \SI{33}{\km} in the transmission fiber span.}
        \label{fig:mdl_span}
    \end{figure}

    Next, the \gls{MDL} estimation correction factor is verified by sweeping the \gls{MDL} and \gls{SNR}. The \gls{MDL} estimation error is defined as the difference between the estimated \gls{MDL} in the setup without noise loading (with \glspl{OSNR} of \SI{40.1}{\dB} and \SI{38.4}{\dB} for the transmitter and in-span emulation schemes, respectively) and the estimated \gls{MDL} with noise loading. \cref{fig:mdl_wo_corr_tx,fig:mdl_wo_corr_span} show the obtained \gls{MDL} estimation error for both emulation schemes. The estimated \gls{MDL} without noise loading is assumed to be the true system \gls{MDL}, as for high \gls{SNR}, according to \eqref{Eq:Lambdarelation}, $\lambda^{2}_{i}\approx\lambda^{2}_{i_{\mathrm{MMSE}}}$. As can be seen from these figures, the \gls{MDL} estimation error increases for low \gls{SNR}, which is expected from \eqref{Eq:Lambdarelation}. A maximum error of up to \SI{5}{\dB} is seen, indicating an underestimation of the system \gls{MDL}.
    
  The eigenvalues used to calculate the \gls{MDL} in \cref{fig:mdl_wo_corr_tx,fig:mdl_wo_corr_span} are now corrected by the correction factor given in \eqref{Eq:roots} and the resulting \gls{MDL} estimation error is given in \cref{fig:mdl_w_corr_tx,fig:mdl_w_corr_span} for the transmitter and in-span \gls{MDL} emulation scheme, respectively. It is seen that the \gls{MDL} error is reduced to below \SI{0.5}{\dB} for the transmitter emulation scheme and to below \SI{0.65}{\dB} for the in-span emulation scheme. For \glspl{SNR} below \SI{14}{\dB}, for both schemes, a small negative \gls{MDL} estimation error is seen, resulting in a overestimation of the system \gls{MDL}.


    \section{Conclusions}\label{sec:conclusions}
    We have experimentally demonstrated an \gls{MDL} estimation technique employing a correction factor that removes the estimation dependence on the \gls{SNR}. When using an \gls{MDL} emulator at the transmitter or an \gls{MDL} emulator placed in the fiber span, the \gls{MDL} estimation correction factor reduced the estimation error, indicating the ability of the technique to improve the \gls{MDL} estimation based on \gls{MMSE} \gls{MIMO} equalizers.

    {\vspace{1mm}\footnotesize \setstretch{1.1} This work was partially supported by the TU/e-KPN Smart Two project, by FAPESP under grants 2018/25414-6, 2017/25537-8, 2015/24341-7, 2015/24517-8, and by the Dutch NWO Gravitation Program on Research Center for Integrated Nanophotonics (Ga 024.002.033).\par}

    \printbibliography

\end{document}